\documentclass[%
 aip,
 amsmath,amssymb,
 reprint,%
]{revtex4-1}

\usepackage{graphicx}
\usepackage{dcolumn}
\usepackage{bm}

\usepackage[utf8]{inputenc}
\usepackage[T1]{fontenc}
\usepackage{mathptmx}
\usepackage{etoolbox}

\makeatletter
\def\@email#1#2{%
 \endgroup
 \patchcmd{\titleblock@produce}
  {\frontmatter@RRAPformat}
  {\frontmatter@RRAPformat{\produce@RRAP{*#1\href{mailto:#2}{#2}}}\frontmatter@RRAPformat}
  {}{}
}%
\makeatother
\begin{document}

\preprint{AIP/123-QED}

\title[Unifying Measurement Schemes in 2-D Terahertz Spectroscopy]{Unifying Measurement Schemes in 2-D Terahertz Spectroscopy}
\author{A. Liu}
\email{aliu1@bnl.gov}
\affiliation{ 
Condensed Matter Physics and Materials Science Division, Brookhaven National Laboratory, Upton, New York, 11973 USA
}%


\begin{abstract}
Two distinct measurement schemes have emerged for the new technique of two-dimensional terahertz spectroscopy (2DTS), complicating the literature. Here, we argue that the `conventional' measurement scheme derived from nuclear magnetic resonance and its optical-frequency analogues should be favored over the `alternative' measurement scheme implemented in the majority of 2DTS literature. It is shown that the conventional scheme avoids issues such as overlapping nonlinearities and facilitates physical interpretation of spectra, in contrast to the alternative scheme.
\end{abstract}

\maketitle

Multidimensional coherent spectroscopy (MDCS) \cite{Cundiff2013} is now a well-known technique with established experimental acquisition protocols in both nuclear magnetic resonance (NMR) \cite{Keeler2010} and at optical frequencies \cite{HammZanni2012,MDCS_Book}. By contrast, its terahertz analogue, termed two-dimensional terahertz spectroscopy (2DTS) \cite{Liu_2DTSReview}, still remains in its infancy with a variety of experimental implementations and measurement schemes. Two distinct measurement schemes have been implemented for 2DTS in the literature. One scheme derives from multidimensional optical spectroscopy and NMR \cite{HammZanni2012,MDCS_Book}, which we refer to as the `conventional scheme'. The other scheme is unique to 2DTS and appears to be introduced by Kuehn et al. \cite{Kuehn2009}, which we refer to as the `alternative scheme'. In this Perspective we argue that the conventional scheme avoids overlapping nonlinearities and facilitates physical interpretation of spectra, and should thus be favored over the alternative scheme.

\begin{figure}[b]
    \centering
    \includegraphics[width=0.4\textwidth]{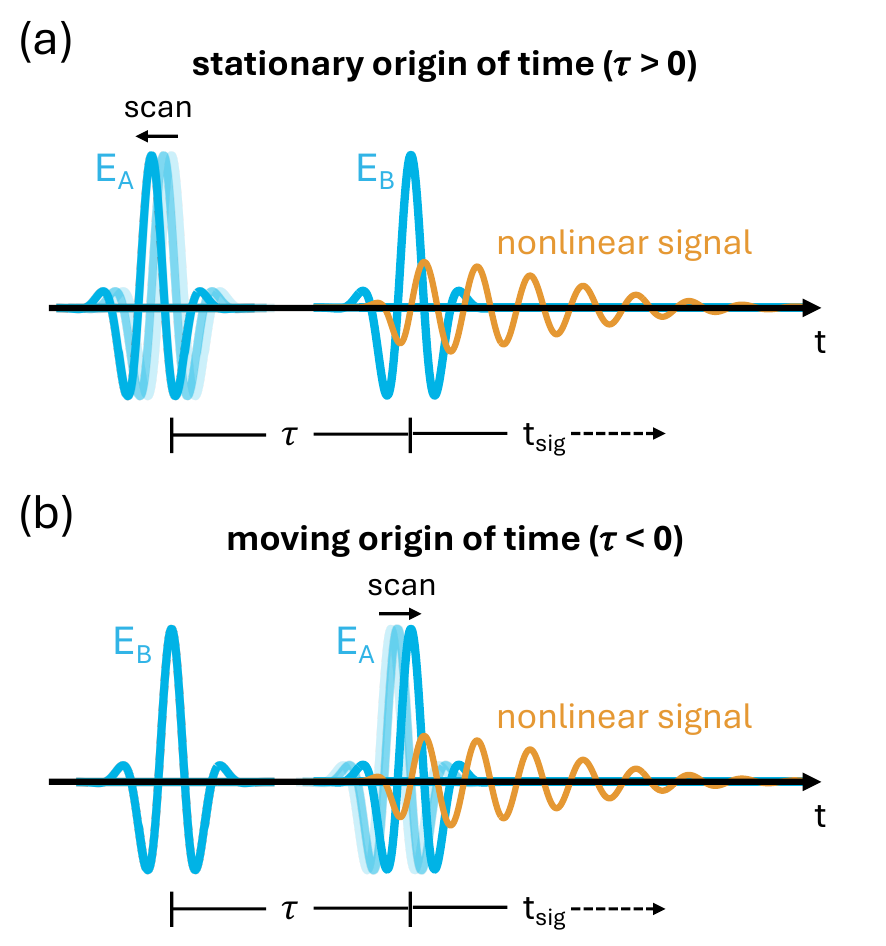}
    \caption{Two conventions for measuring nonlinear signal emission. (a) For $E_A$ arriving before $E_B$, $\tau$ is defined to be positive and the origin of time $t_{sig}$ is stationary. (b) For $E_B$ arriving before $E_A$, $\tau$ is defined to be negative and the origin of time $t_{sig}$ moves with changing $\tau$.}
    \label{Fig1}
\end{figure}

We first define the excitation protocol used in most implementations of 2DTS, comprised of two pulses $E_A$ and $E_B$. In anticipation of defining the two measurement schemes, the inter-pulse time delay $\tau$ is defined to be (positive)negative when ($E_A$)$E_B$ arrives first, and $E_B$ is always stationary in time. The time axis of nonlinear signal emission $t_{sig}$ is then defined by setting its origin to the arrival of the second pulse. We now find distinct scenarios for the two excitation pulse time-orderings. For positive inter-pulse delay ($\tau > 0$, $E_A$ arrives before $E_B$), the origin of time $t_{sig} = 0$ is stationary and defined by the fixed pulse $(E_B)$ as shown in Fig.~\ref{Fig1}(a). For negative inter-pulse delay ($\tau < 0$, $E_B$ arrives before $E_A$), $t_{sig} = 0$ advances in time and is defined by the moving pulse $(E_A)$ as shown in Fig.~\ref{Fig1}(b). 

With the two excitation pulse time-orderings in mind, we must now distinguish between the time axis of the nonlinear signal emission $t_{sig}$ (whose origin is defined by the arrival of the final excitation pulse) and the laboratory observation time $t$ (along which we Fourier transform). In the conventional scenario depicted in Fig.~\ref{Fig1}(a), the two time variables $t_{sig}$ and $t$ both have a stationary origin of time and can be taken to be equivalent. In the second scenario depicted in Fig.~\ref{Fig1}(b), however, the two variables are related by a transformation $t = t_{sig} - \tau$ (recall that $\tau$ is negative for this excitation time-ordering). Below, we examine the consequences of this difference in 2DTS spectra.

\begin{figure}[t]
    \centering
    \includegraphics[width=0.5\textwidth]{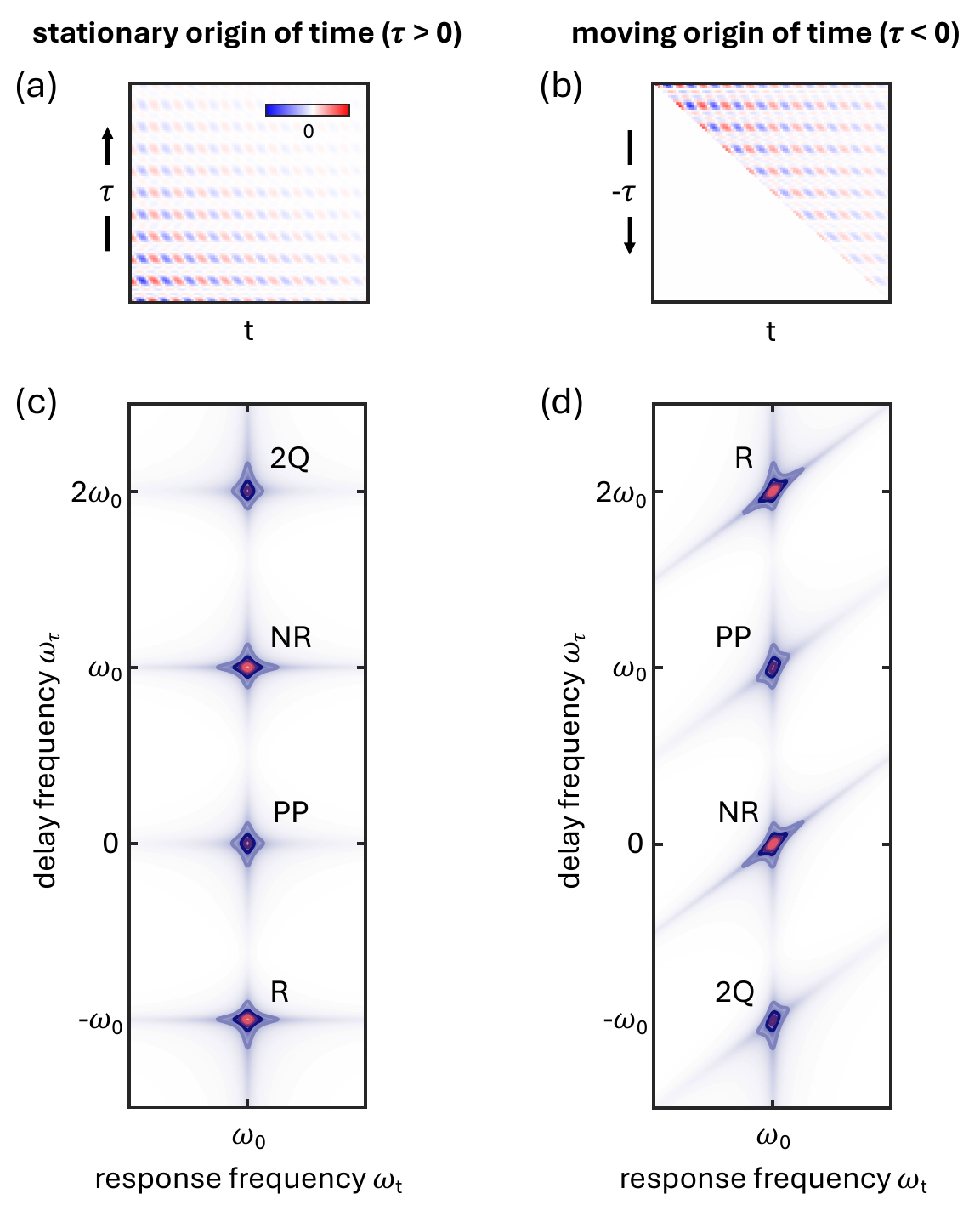}
    \caption{Time-domain nonlinear signal for a (a) stationary and (b) moving origin of time, corresponding to the excitation time-orderings shown in Figs.~\ref{Fig1}(a) and \ref{Fig1}(b) respectively. Corresponding amplitude 2DTS spectra are shown in Figs.~\ref{Fig2}(c) and \ref{Fig2}(d) respectively, which exhibit drastically different peak positions for each nonlinearity. Dephasing limited by population relaxation ($T_2 = 2T_1$) was assumed.}
    \label{Fig2}
\end{figure}

For the two scenarios of positive and negative inter-pulse delay described above, optical responses deriving from physically-identical nonlinearities appear at different points in frequency-space. We demonstrate this schematically for the third-order nonlinearities of a quantum-ladder system (in which the dephasing time $T_2$ is limited by the population relaxation time $T_1$ for simplicity), whose nonlinear signal is plotted in the time-domain for a stationary and moving origin of time in Figs.~\ref{Fig2}(a) and \ref{Fig2}(b) respectively. Note that the horizontal axis $t$ is the laboratory time with a stationary origin, causing the origin of $t_{sig}$ to move with a change in $\tau$.

\begin{figure*}
    \centering
    \includegraphics[width=0.6\textwidth]{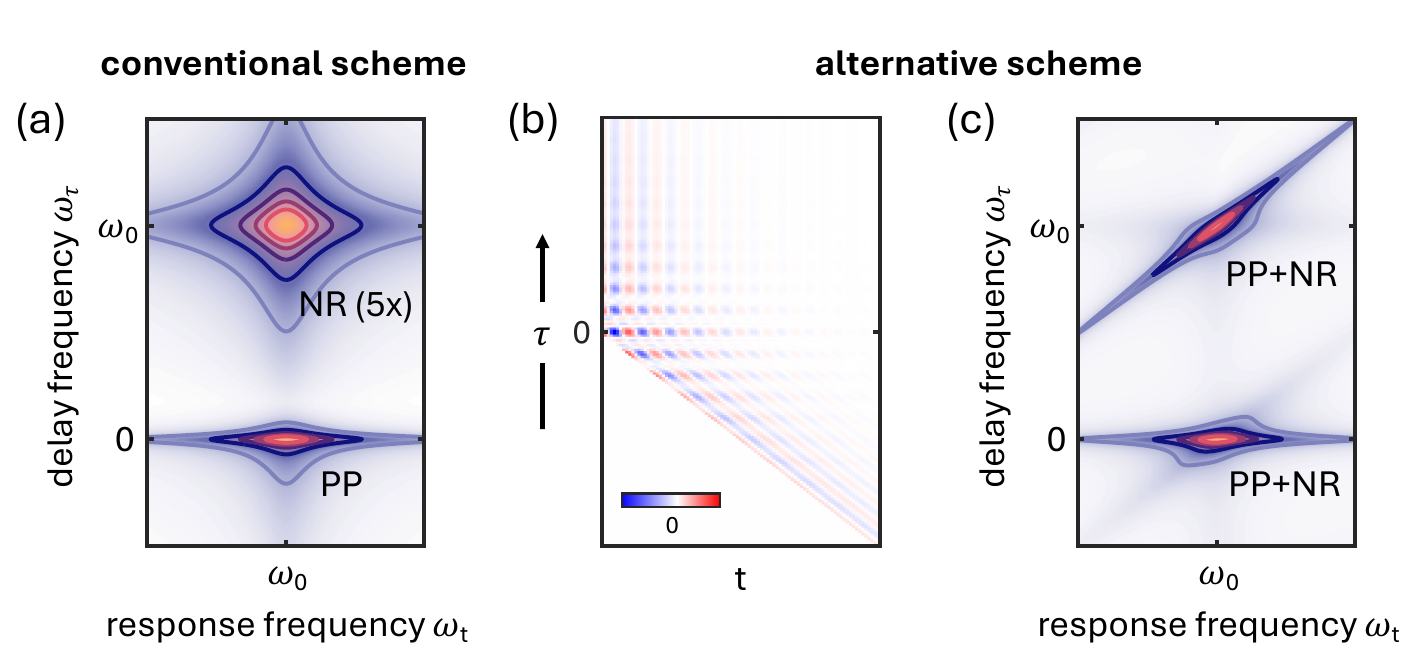}
    \caption{Nonlinear signal for the typical situation of dominant pure dephasing ($T_2 \ll T_1$). (a) 2DTS spectrum acquired in the conventional scheme which separates the non-rephasing and pump-probe nonlinearities and their respective dephasing and relaxation timescales. Amplitude of the non-rephasing nonlinearity was scaled by a factor of 5 to show the lineshape difference. (b) Time-domain signal acquired in the alternative scheme with both excitation time-orderings. (c) Corresponding amplitude 2DTS spectrum, in which both peaks at $\omega_\tau = 0$ and $\omega_\tau = \omega_0$ are dominated by the pump-probe nonlinearity and the dephasing and relaxation timescales are intertwined.}
    \label{Fig3}
\end{figure*}

We then Fourier transform the time-domain signals into the frequency-domain to obtain their respective 2DTS spectra shown in Figs.~\ref{Fig2}(c) and \ref{Fig2}(d). Four third-order nonlinearities are observed in both spectra that evolve with a response frequency $\omega_t = \omega_0$, termed in the literature \cite{HammZanni2012,MDCS_Book} as rephasing (R), pump-probe (PP), non-rephasing (NR), and two-quantum (2Q). Between the two spectra, the various nonlinearities manifest at drastically different positions along the delay frequency $\omega_\tau$ due to the opposite sign of inter-pulse delay and the rotating frame imposed by the transformation $t = t_{sig} - \tau$ described above. 

Having now clarified how various third-order nonlinearities appear in 2DTS spectra for either a stationary or moving origin of time $t_{sig}$, we can now define the two measurement schemes that form the crux of this Perspective. The `conventional scheme' is directly shown in Figs.~\ref{Fig2}(a) and \ref{Fig2}(c), where only a single excitation time-ordering is measured and the origin of time $t_{sig}$ is stationary. Applications of this conventional scheme in the 2DTS literature have been limited \cite{Lu2016,Johnson2019,Zhang2024,Liu_2023_echo,Barbalas2025}. The `alternative scheme' involves simultaneous measurement of both excitation time-orderings shown in Figs.~\ref{Fig2}(a) and \ref{Fig2}(b), and the resultant 2DTS spectra is thus a combination of both spectra shown in Figs.~\ref{Fig2}(c) and \ref{Fig2}(d). Applications of this alternative scheme in the literature have been more widespread \cite{Somma2016,Mahmood2021,Pal2021,Lin2022,Blank2023,Luo2023,Kim2024,Katsumi2024}. In the remainder of this Perspective, we advocate for the conventional scheme over the alternative scheme with two primary arguments: (1) We first show that the conventional scheme cleanly separates distinct third-order nonlinearities which overlap in the alternative scheme. (2) We then demonstrate that peak positions and patterns in the conventional scheme facilitate direct physical interpretation, which is obfuscated in the alternative scheme.

To demonstrate how overlapping nonlinearities in the alternative scheme can lead to ambiguities of the nonlinear optical response, we again examine the third-order nonlinearities of a quantum-ladder system. Now we assume that the system experiences significant pure dephasing of coherences, resulting in a significant difference between dephasing and population relaxation times ($T_2 \ll T_1$). The 2DTS spectrum of an identical system acquired in the conventional scheme is shown in Fig.~\ref{Fig3}(a), in which the non-rephasing and pump-probe nonlinearities are cleanly separated. Here, the large difference between dephasing and population relaxation timescales is obvious by inspection, and can be extracted without ambiguity by conventional methods \cite{Siemens2010} in all cases. For comparison, the time-domain nonlinear signal is shown in Fig.~\ref{Fig3}(b) for the alternative measurement scheme, in which both excitation time-orderings are measured simultaneously. The corresponding 2DTS spectrum is shown in Fig.~\ref{Fig3}(c), in which both peaks at $\omega_\tau = 0$ and $\omega_\tau = \omega_0$ result from a combination of the pump-probe and non-rephasing nonlinearities and exhibit identical characteristic linewidths. Furthermore, a contribution of two distinct nonlinearities and a moving reference frame results in spectral lineshapes that, besides the limiting cases of dominant population relaxation and dominant dephasing, cannot be fitted with the usual analytical expressions \cite{Siemens2010}. Note that the rephasing nonlinearity (not shown in Fig.~\ref{Fig3}) will remain isolated in the absence of a two-quantum nonlinearity, allowing for extraction of the dephasing time $T_2$, but will suffer from the same issue described above if this condition is not met. Other physics such as inhomogeneous broadening and excitation-induced effects \cite{MDCS_Book} can even further complicate the analysis. 

\begin{figure}[b]
    \centering
    \includegraphics[width=0.5\textwidth]{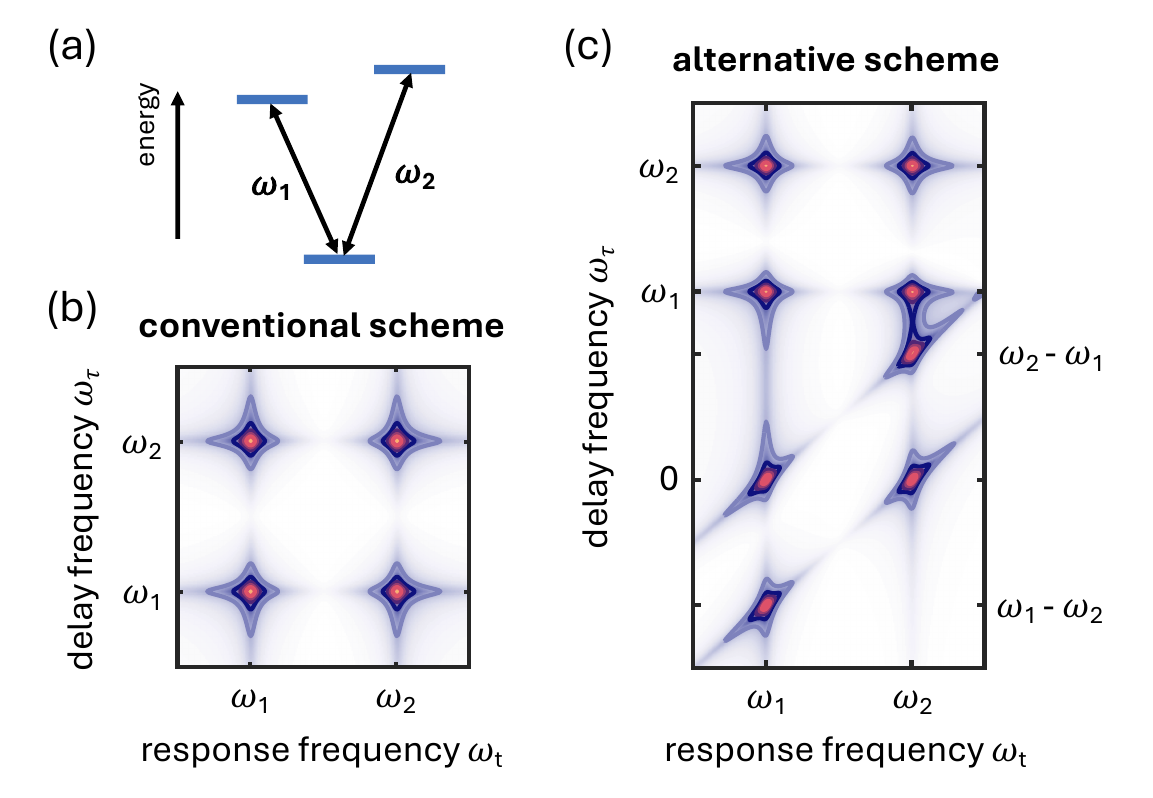}
    \caption{(a) Quantum `vee' level system, consisting of two transitions of frequencies $\omega_1$ and $\omega_2$ as indicated. Identical linewidths and transition dipole moments are assumed between the transitions for simplicity. (b,c) Corresponding amplitude 2DTS spectra, with only non-rephasing nonlinearities shown, are plotted for the (b) conventional scheme and (c) alternative scheme. The simultaneous measurement of two time-orderings doubles the number of non-rephasing peaks in (c) with respect to the four peaks in (b).}
    \label{Fig4}
\end{figure}

While the above comparisons already hint that the alternative measurement scheme muddles physical interpretation of peaks in 2DTS spectra, the advantage of the conventional measurement scheme becomes obvious once we consider a system that involves coupling. Here, we consider a quantum `vee' level system shown in Fig.~\ref{Fig4}(a), consisting of two optical resonances frequencies $\omega_1$ and $\omega_2$. These two transitions are coupled through a shared ground state, which is found straightforwardly in Fig.~\ref{Fig4}(b) by a 2DTS spectrum acquired in the conventional scheme. For the non-rephasing nonlinearities shown, the two `on-diagonal' peaks located at $(\omega_\tau,\omega_t) = (\omega_1,\omega_1)$ and $(\omega_\tau,\omega_t) = (\omega_2,\omega_2)$ inform the two optical transitions, and the two `off-diagonal' peaks located at $(\omega_\tau,\omega_t) = (\omega_1,\omega_2)$ and $(\omega_\tau,\omega_t) = (\omega_2,\omega_1)$ indicate that they are coherently coupled.

In contrast, the 2DTS spectrum acquired with the alternative scheme is far more complicated. For simplicity we only show the non-rephasing nonlinearities, which are now doubled in number. Additional peaks now appear at $\omega_\tau = \{\omega_1-\omega_2,0,\omega_2-\omega_1\}$, which can lead to incorrect interpretations of additional resonances and coupling thereof. Including the other rephasing, pump-probe, and two-quantum nonlinearities for complex systems often renders physical interpretation of spectra unfeasible altogether.

We now discuss potential reasons for favoring the alternative scheme over the conventional scheme. A scenario typically cited to favor the alternative scheme is when a nonlinear signal is only generated during overlap of the excitation pulses, leading to the origin of $\tau$ being poorly defined. Whether due to short system decay times or off-resonant excitation, however, we have previously shown \cite{Liu2024_OptExpress} that both such cases lead to 2DTS spectra whose lineshapes are largely defined by the excitation spectrum. Therefore in both cases a 2DTS measurement (with both the conventional and alternative measurement schemes) provides minimal advantage over a simple one-dimensional measurement of the nonlinear signal along either $\tau$ or $t$, which informs the strength of the optical nonlinearity.

Another common reason cited in favor of the alternative scheme is to reduce oscillatory artifacts in 2DTS spectra that result from incomplete sampling of the time-domain signal along delay $\tau$. While truncating the nonlinear signal reduces the necessary data to be acquired, this advantage is usually outweighed by the additional redundant (for a collinear geometry) data measured for the second time-ordering in the alternative scheme. We finally note that fully sampled nonlinear signals in the conventional and alternative schemes will yield spectra of identical spectral resolution (despite the latter being composed of twice the number of sampling points) after enforcing causality with zero-padding \cite{Bartholdi1973} in the former scheme.

In this Perspective, we have compared the two primary measurement schemes employed in the 2DTS literature and clarified various disadvantages of the `alternative' scheme used by most authors. The disadvantages of overlapping nonlinearities and more complicated interpretation were demonstrated for nonlinearities of a quantum-ladder system, but analogous considerations apply to 2DTS of a classical nonlinear oscillator \cite{Liu2024_OptExpress} as well. We also emphasize that in a collinear geometry with identical excitation pulses, employed in most studies, measuring with the alternative scheme results in superfluous information since the different time-orderings generate redundant signals. The issues discussed here could be remedied to some degree by phase-matching in a non-collinear excitation geometry \cite{Liu_2023_echo}, but we advocate for the conventional scheme and unifying measurement schemes in 2DTS with their established counterparts across the electromagnetic spectrum. 

\begin{acknowledgments}
    Albert Liu was supported by the U.S. Department of Energy, Office of Basic Energy Sciences, under Contract No. DE-SC0012704.
\end{acknowledgments}

\section*{Data Availability Statement}

Data sharing is not applicable to this article as no new data were created or analyzed in this study.

\nocite{*}
%

\end{document}